\documentclass[
prb,
superscriptaddress,
amsmath, amssymb,
amssymb,
reprint,
floatfix
]{revtex4-1}

\usepackage{graphicx}
\usepackage{physics} 
\usepackage{bm}
\usepackage{color}
\usepackage[]{natbib}
\usepackage{amsmath,amssymb,amsfonts}
\usepackage[normalem]{ulem}
\usepackage{hyperref}	

\begin{document}

\title{A Spin Quintet in a Silicon Double Quantum Dot: Spin Blockade and Relaxation}

\author{Theodor Lundberg} \thanks{twl28@cam.ac.uk}
\affiliation{Cavendish Laboratory, University of Cambridge, J.J. Thomson Avenue, Cambridge CB3 0HE, UK}
\affiliation{Hitachi Cambridge Laboratory, J.J. Thomson Avenue, Cambridge CB3 0HE, UK}
\author{Jing Li}
\affiliation{Univ. Grenoble Alpes, CEA, IRIG, MEM/L\_Sim, 38000 Grenoble, France}
\author{Louis Hutin}
\affiliation{CEA/LETI-MINATEC, CEA-Grenoble, 38000 Grenoble, France}
\author{Benoit Bertrand}
\affiliation{CEA/LETI-MINATEC, CEA-Grenoble, 38000 Grenoble, France}
\author{David J. Ibberson}
\affiliation{Quantum Engineering Technology Labs, University of Bristol, Tyndall Avenue, Bristol BS8 1FD, UK}
\affiliation{Hitachi Cambridge Laboratory, J.J. Thomson Avenue, Cambridge CB3 0HE, UK}
\author{Chang-Min Lee}
\affiliation{Department of Materials Science and Metallurgy, University of Cambridge, 27 Charles Babbage Road, Cambridge CB3 0FS, UK}
\author{David J. Niegemann}
\affiliation{CNRS, Grenoble INP, Institut N\'{e}el, University of Grenoble Alpes, 38000 Grenoble, France}
\author{Matias Urdampilleta}
\affiliation{CNRS, Grenoble INP, Institut N\'{e}el, University of Grenoble Alpes, 38000 Grenoble, France}
\author{Nadia Stelmashenko}
\affiliation{Department of Materials Science and Metallurgy, University of Cambridge, 27 Charles Babbage Road, Cambridge CB3 0FS, UK}
\author{Tristan Meunier}
\affiliation{CNRS, Grenoble INP, Institut N\'{e}el, University of Grenoble Alpes, 38000 Grenoble, France}
\author{Jason W. A. Robinson}
\affiliation{Department of Materials Science and Metallurgy, University of Cambridge, 27 Charles Babbage Road, Cambridge CB3 0FS, UK}
\author{Lisa Ibberson}
\affiliation{Hitachi Cambridge Laboratory, J.J. Thomson Avenue, Cambridge CB3 0HE, UK}
\author{Maud Vinet}
\affiliation{CEA/LETI-MINATEC, CEA-Grenoble, 38000 Grenoble, France}
\author{Yann-Michel Niquet}
\affiliation{Univ. Grenoble Alpes, CEA, IRIG, MEM/L\_Sim, 38000 Grenoble, France}
\author{M. Fernando Gonzalez-Zalba} \thanks{mg507@cam.ac.uk}
\affiliation{Hitachi Cambridge Laboratory, J.J. Thomson Avenue, Cambridge CB3 0HE, UK}

\begin{abstract}
Spins in gate-defined silicon quantum dots are promising candidates for implementing large-scale quantum computing. To read the spin state of these qubits, the mechanism that has provided the highest fidelity is spin-to-charge conversion via singlet-triplet spin blockade, which can be detected in-situ using gate-based dispersive sensing. In systems with a complex energy spectrum, like silicon quantum dots, accurately identifying when singlet-triplet blockade occurs is hence of major importance for scalable qubit readout. In this work, we present a description of spin blockade physics in a tunnel-coupled silicon double quantum dot defined in the corners of a split-gate transistor. Using gate-based magnetospectroscopy, we report successive steps of spin blockade and spin blockade lifting involving spin states with total spin angular momentum up to $S=3$. More particularly, we report the formation of a hybridized spin quintet state and show triplet-quintet and quintet-septet spin blockade. This enables studies of the quintet relaxation dynamics from which we find $T_1 \sim 4 ~\mu s$. Finally, we develop a quantum capacitance model that can be applied generally to reconstruct the energy spectrum of a double quantum dot including the spin-dependent tunnel couplings and the energy splitting between different spin manifolds. Our results open for the possibility of using Si CMOS quantum dots as a tuneable platform for studying high-spin systems. 
\end{abstract}

\maketitle


High-spin states have been shown to play a key role in a variety of important physical phenomena. They are involved in singlet fission in organic photovoltaics \cite{tayebjee2017quintet,smith2010singlet}, unconventional high-spin superconductivity \cite{kim2018beyond,yu2018singlet}, and the energy states of molecules and complexes with large delocalised electron systems that are relevant, for example, to biochemical catalysis \cite{swart2004validation,chapyshev2000quintet, teki2001pi}. Similarly, spin-based quantum computing also necessitates a comprehensive understanding of the interplay of the various spin states that exist in the quantum computing platform of choice \cite{malinowski2018spin, chen2017spin, van2018readout, leon2019coherent}. Spins bound to quantum dots defined in silicon have garnered significant attention as an attractive quantum computing platform due to their long coherence times \cite{tyryshkin2012electron,saeedi2013room}, the compatibility with industrial manufacturing techniques \cite{maurand2016cmos, veldhorst2017silicon, crippa2019gate}, and recently also due to the demonstration of high fidelity one- and two-qubit operations \cite{yoneda2018quantum, huang2019fidelity, watson2018programmable}. In spin qubits, quantum state readout is achieved via spin-to-charge conversion which translates the spin state into a selective movement of charge that can be efficiently detected using charge sensors~\cite{morello2010single} or resonant circuits~\cite{schroer2012radio}. Currently, the spin-to-charge mechanism that has enabled highest fidelity readout, even at high temperatures and moderate magnetic fields~\cite{yang2019silicon_one_kelvin, zhao2018coherent}, is Pauli spin blockade of the two-electron singlet and triplet states \cite{kloeffel2013prospects, fogarty2018integrated, harvey2018high}. Under spin blockade, electron transitions from one quantum dot to another are prohibited in the triplet manifold due to the Pauli exclusion principle, while transitions among singlet states are permitted. However, singlet-triplet spin blockade can be lifted by the presence of low-energy excited states that enable direct triplet tunneling and render spin blockade less effective~\cite{betz2015dispersively}. Given that silicon possesses an additional valley degree of freedom\cite{culcer2012valley}, the energy spectrum in silicon quantum dots can be rather complex and accurately identifying when singlet-triplet blockade occurs is hence of importance for achieving reliable and scalable readout of spin qubits in silicon. 

\begin{figure*}
\centering
\includegraphics[width=\textwidth]{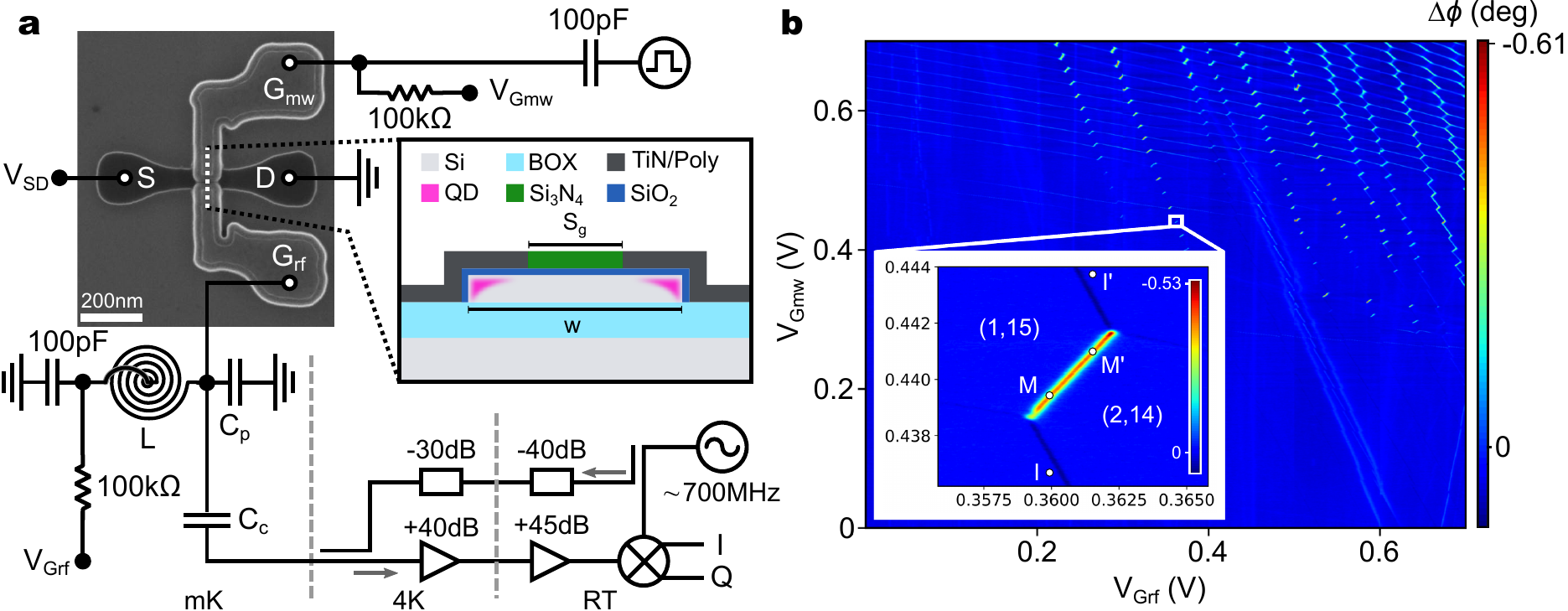}
\caption{Measurement setup and charge stability diagram of the DQD. \textbf{(a)} Scanning electron microscopy (SEM) micrograph and sketch of the transistor and measurement setup. The SEM micrograph is taken during processing, after the definition of the Si$_3$N$_4$ spacers. The sketch inset capturing the cross-section of the transistor perpendicular to the channel illustrates the architecture of the device as well as the double quantum dot (DQD) formed in the channel corners when applying a positive potential to the split-gates. Gate G$_{\text{rf}}$ is connected to a resonant LC circuit in which the inductive component is a superconducting NbN planar spiral inductor. \textbf{(b)} Charge stability diagram recorded using gate-based dispersive readout while sweeping the potentials of G$_{\text{rf}}$ and G$_{\text{mw}}$, showing regular features characteristic to a DQD. Counting the number of charge transitions in the diagram allows identification of the exact DQD charge configurations, e.g. that shown in the inset which highlights the (1,15)-(2,14) interdot charge transition investigated in this letter. The points I, I$^\prime$, M and M$^\prime$ indicate the initialisation and measurement points of pulsing experiments presented in Figure 3.}
\label{fig1}
\end{figure*}

In this letter, we go beyond the standard description of singlet-triplet Pauli spin blockade and demonstrate that low lying excited states can lead to successive steps of spin blockade and spin blockade lifting which span spin states with total spin angular momentum up to $S=3$. We do so using a double quantum dot (DQD) defined electrostatically in the corners of a Si complementary metal-oxide-semiconductor (CMOS) split-gate transistor~\cite{betz2015dispersively}. By using gate-based dispersive readout~\cite{petersson2010charge,colless2013dispersive,ahmed2018radio} and magnetospectroscopy of an interdot charge transition (ICT) between two quantum dots containing a total of 2 and 14 electrons respectively, we demonstrate the formation of a hybridized spin quintet state ($S=2$) between the quantum dots. We show how spin quintet tunnelling can be blocked at low magnetic fields by the triplet states ($S=1$) and at high fields by a spin septet state ($S=3$). The spin blockade is used to study the quintet spin relaxation to the triplet and septet states, which we find to be of the order of a few microseconds. Moreover, to better understand the magnetic dependence of the dispersive response, we develop a quantum capacitance model that enables reconstruction of the energy spectrum of the coupled DQD, including the spin-dependent tunnel coupling and the energy splitting between different spin manifolds. Overall, our study provides a comprehensive understanding of spin blockade physics in systems with a dense energy spectrum and opens for the possibility of investigating the dynamics of high-spin systems using programmable CMOS technology.

Figure 1a shows a scanning electron microscopy (SEM) micrograph of an $n$-type CMOS split-gate transistor fabricated on a 300 mm silicon-on-insulator (SOI) wafer similar to the device presented here. The inset of Figure 1a presents the cross-section of the silicon channel of the transistor, which has a height $h = 7~\text{nm}$, a width $w = 70~\text{nm}$ and is separated from the Si substrate, which we hold at $0 ~\text{V}$, by a 145~nm SiO$_2$ buried oxide (BOX). On top and separated from the channel by $6~\text{nm}$ of SiO$_2$, there is a pair of face-to-face gate electrodes (G$_{\text{rf}}$ and G$_{\text{mw}}$), which have gate length $L_g = 60~\text{nm}$ and are split from each other by $S_g = 40~\text{nm}$. The space between and around the gates is covered by 34~nm-wide Si$_3$N$_4$ spacers which help separate the highly-$n$-type-doped source (S) and drain (D) contacts from the central part of the intrinsic channel. The channel region below G$_{\text{rf}}$ is lighty Bi doped with an average of $\sim 1$ Bi dopant under the area of the G$_{\text{rf}}$ gate. By increasing the electrostatic potential of G$_{\text{rf}}$ and G$_{\text{mw}}$, electrons accumulate in the corners of the channel thus forming two quantum dots, QD$_{\text{rf}}$ and QD$_{\text{mw}}$, in parallel between the source and drain, which have controllable electron occupancies (see inset of Figure 1a)~\cite{voisin2015electrical}. Electrons can be drawn in to the quantum dots from the S and D reservoirs, which we hold at $0 ~\text{V}$. 

We control and read the electron configuration of the DQD via the setup presented in Figure 1a. The microwave gate (G$_{\text{mw}}$) is connected to a DC line ($V_\text{Gmw}$) and a high frequency line through a bias tee. The other gate, the reflectometry gate (G$_{\text{rf}}$), is connected to a DC line ($V_\text{Grf}$) and, in parallel, to a lumped-element LC resonant circuit comprised of a superconducting NbN planar spiral inductor as well as the parasitic capacitance to ground, $C_\text{p}$. At $B = 0~\text{T}$, the resonance frequency of the resonator is $f_0 = 704.68 ~\text{MHz}$; however when increasing the magnetic field, the kinetic inductance of the superconducting spiral inductor increases, leading to shifts in $f_0$ according to $f_0(B) = 1 / (2\pi \sqrt{L(B)(C_{p}+C_\text{c}+C_\text{d})})$, where $L(B)$ is the magnetic field dependent inductance of the spiral inductor, $C_\text{c}=0.2~\text{pF}$ is the coupling capacitance and $C_\text{d}$ is the state-dependent device capacitance. We drive the resonator close to resonance with a frequency $f_{\text{rf}}$ and power $P_{\text{rf}}$ at the input of the resonator in the range of $-105$ to $-95 ~\text{dBm}$. The signal reflected from the resonant circuit is amplified by $40 ~\text{dB}$ at $4~\text{K}$ and by further $45 ~\text{dB}$ at room temperature, where it is subsequently homodyne IQ demodulated and low-pass filtered, thus allowing measurement of the phase shift of the reflected signal, $\Delta \phi$. Variations in phase, $\Delta\phi=-2Q_\text{l}\Delta C_d/(C_{p}+C_\text{c}+C_\text{d})$ where $Q_l$ is the loaded quality factor, arise due to changes in device capacitance that occur, for example, when single electrons tunnel cyclically between QDs, or between a QD and a reservoir, because of the influence of the rf drive~\cite{mizuta2017quantum}. 

\begin{figure*}
\centering
\includegraphics[width=\textwidth]{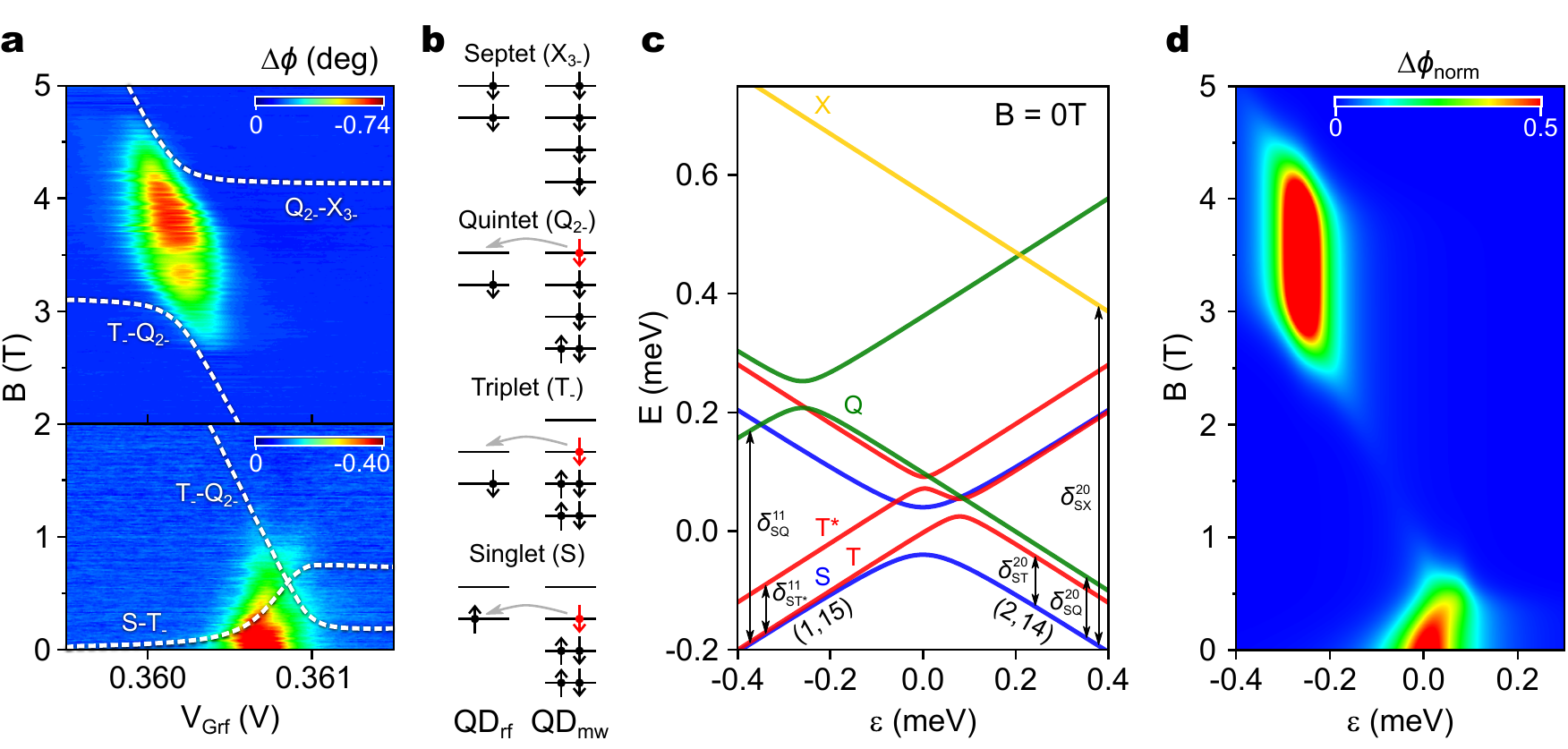}
\caption{Pauli spin blockade and energy levels of spin states with $S\leq3$. \textbf{(a)} Measurement of the (1,15)-(2,14) interdot charge transition at $V_\text{Gmw} = 0.440~\text{V}$ as a function of external in-plane magnetic field. From bottom to top, the dashed lines indicate fits of the S-T$_-$, T$_-$-Q$_{2-}$ and Q$_{2-}$-X$_{3-}$ intersection points at varying magnetic field. \textbf{(b)} Illustrative single particle energy diagrams of the singlet S ($m_s=0$), triplet T$_-$ ($m_s=-1$), quintet Q$_{2-}$ ($m_s=-2$), and septet X$_{3-}$ ($m_s=-3$) states that successively become the ground state when increasing the magnetic field in (a). For simplicity, the energy diagrams omit the five lowest lying energy levels of QD$_\text{mw}$. The (1,15)-(2,14) anti-crossings of the singlet, triplet, and quintet shown in (c) are demonstrated by an electron (in red) tunneling from QD$_\text{mw}$ to QD$_\text{rf}$. \textbf{(c)} Energy levels of the double quantum dot simulated based on parameters extracted from (a) as a function of (1,15)-(2,14) detuning. The different colours indicate states of varying multiplicity and spin angular momentum S: a singlet with $S = 0$ (blue), triplets with $S = 1$ (red), quintets with $S = 2$ (green) and septets with $S = 3$ (yellow). The lines with positive slope are (1,15) states, while lines with negative slope are (2,14) states. These states mix near the anti-crossings due to (1,15)-(2,14) interdot tunneling. \textbf{(d)} Normalised phase response as a function of magnetic field simulated based on the energy levels presented in (c). In order to enhance the visibility of low-field features, the z-scale is limited to half of the normalised phase response.}
\label{fig2}
\end{figure*}

We present the charge stability diagram in Figure 1b\footnote{The charge stability diagram in Figure 1b is stitched together by 49 individual diagrams, each of dimension 0.1 V by 0.1 V} obtained by measuring the phase response as a function of the voltages $V_\text{Grf}$ and $V_\text{Gmw}$. The hexagon-like features in the diagram confirm the presence of a DQD in the Si channel and the lines of non-zero phase shift indicate regions of charge bistability. The short lines with positive slope indicate electron tunneling between dots whereas the quasi-vertical (quasi-horizontal) lines correspond to regions where QD$_{\text{rf}}$ (QD$_{\text{mw}}$) exchanges an electron with the source/drain reservoirs. We note a larger voltage period of the lines in $V_\text{Grf}$ with respect to $V_\text{Gmw}$, which we attribute to a misalignment of $7 \pm 3$~nm in the placement of the gates on the Si channel in which G$_{\text{mw}}$ overlaps the channel more so than G$_{\text{rf}}$. The increased elongation and the rounding of the corners of the hexagons at larger gate voltages indicate an increased tunnel coupling between the dots \cite{van2002electron}. Gate-based reflectometry readout enables us to count the number of electrons down to the first ones by counting voltage shifts in the vertical and horizontal lines (see Supporting Information S1). The inset of Figure 1b shows the region of the charge stability diagram around the even-parity interdot charge transition (ICT) between the (1,15) and (2,14) DQD charge configurations. Here $i$ in $(i,j)$ refers to the electron occupancy of QD$_{\text{rf}}$ while $j$ refers to the electron occupancy of QD$_{\text{mw}}$. The tunneling between the two dots results in a change of device capacitance given by the quantum capacitance of the system, which in the slow relaxation limit can be described by the expression~\cite{mizuta2017quantum} \begin{equation} C_\text{Q}=-\sum_i e\alpha \frac{\partial^2 E_i}{\partial V_\text{Grf}^2}P_i \end{equation} where $e$ is the electron charge, $\alpha=\alpha_\text{rf}-\alpha_\text{mw}$ is the difference between the capacitive coupling ratio of G$_\text{rf}$ to QD$_{\text{rf}}$ and QD$_{\text{mw}}$, $E_i$ is the eigenenergy of the many-particle state $i$, and $P_i$ is the probability of the state $i$ being occupied. The points I, I$^\prime$, M, and M$^\prime$ in the inset indicate the initialisation and measurement points used for pulsing experiments described later in this letter. In the following, we focus on the region and ICT shown in the inset of Figure 1b.

In order to study the spin physics of the multi-electron DQD defined in this device, and in particular the ICT shown in Figure 1b, we perform a dispersive magnetospectroscopy study by measuring the line-trace intersecting the ICT at $V_\text{Gmw} = 0.440$ V while increasing the magnetic field, $B$, which is applied in-plane with the device and at an $83^{\circ}$ angle to the nanowire. To account for the magnetic field dependent inductance $L(B)$, which changes significantly for $B\gtrsim2~\text{T}$, the magnetospectroscopy study is split into two measurements as presented in Figure 2a, where $f_{\text{rf}}$ is adjusted accordingly for $B=2$ to $5~\text{T}$. The measurements show that the signal seen at $B=0~\text{T}$ in Figure 1b disappears when $B>1.2~\text{T}$, and that a new signal at slightly lower $V_{\text{Grf}}$ appears at $B \sim 2.6~\text{T}$ and eventually disappears again at $B \sim 4.4~\text{T}$. 

The signal at $B = 0~\text{T}$ arises from tunneling between the anti-crossing singlet states S(1,15) and S(2,14) as given by Equation (1) and as sketched in the lower panel of Figure 2b. The curvature of the anti-crossing singlet states is apparent from their calculated eigenenergies that appear in blue in Figure 2c as function of energy detuning between dots $\varepsilon=e\alpha(V_\text{Grf}-V_\text{Grf}^\text{0})$, where $V_\text{Grf}^\text{0}=0.3607$~V in this experiment. When $B$ is increased, excited states with a non-zero spin angular momentum projection onto the direction of $B$, $m_s\neq0$, will Zeeman-split according to $E_Z=m_sg\mu_BB$, where $g$ is the Land\'{e} $g$-factor and $\mu_B$ is the Bohr magneton. A Zeeman-split state, for example the triplet T$_-$ with $m_s=-1$ illustrated in Figure 2b, can therefore become the ground state at sufficiently large $B$. In this scenario, the tunneling between T(1,1)- and S(2,0)-like states -- T$_-$(1,15) and S(2,14) here -- is forbidden by the Pauli exclusion principle~\cite{lai2011pauli, petta2005coherent, maurand2016cmos, west2019gate}, thereby leading to Pauli spin blockade: the ICT signal disappears because T$_-(1,15)$ shows no curvature at detuning $\varepsilon = 0$ and thus makes no contribution to $C_\text{Q}$. One may therefore initially be led to believe that the lack of signal from $B=1.2-2.6~\text{T}$ in Figure 2a is due to singlet-triplet spin blockade and that the signal at $B=2.6-4.4~\text{T}$ arises from the curvature of the T$_-$(1,15)-T$_-$(2,14) anti-crossing \cite{betz2015dispersively, house2015radio, landig2019microwave}. As we explain below, this is however not the case.

In the lower part of Figure 2a, we note that as $B$ is increased from 0 to 0.8~T, the ICT signal decreases in intensity asymmetrically from the left. The reduction of signal can be ascribed to the lowering of T$_-$ below the singlet ground state. This is illustrated by the dashed white line labelled S-T$_-$ in Figure 2a, which tracks the position of the S-T$_-$ crossing above which T$_-$ becomes the new ground state of the system as previously explained (see Supporting Information S2 for energy diagrams similar to Figure 2c at non-zero magnetic fields). Because the signal of the singlet disappears at lower $V_\text{Grf}$ first, the anti-crossing between T$_-$(1,15) and T$_-$(2,14) states must appear at larger gate voltage. For that reason, the signal in the region $B=2.6-4.4~\text{T}$ cannot arise from triplet tunneling but, as we demonstrate below, instead comes from hybridised quintet spin states ($S=2$). By closer inspection of the $B=0.8-1.2~\text{T}$ region (Supporting Information S3), we do however identify the signal of the T$_-$ anti-crossing. This signal overlaps that of the low-field singlet because the singlet-triplet splitting of this system, $\delta_\text{ST}$, is small relative to the tunnel coupling of the singlet states, $\Delta_\text{S}$, thereby making the two signals difficult to discern. The triplet signal is accompanied by a second peak (Supporting Information S3) arising from the tunneling between an excited triplet, T*$_-$(1,15), and T$_-$(2,14) as illustrated with the red eigenenergies in Figure 2c. This additional triplet anti-crossing explains the extension of the low-field ICT signal beyond the S-T$_-$ crossing line.


As we increase the magnetic field further, the triplet signals vanish at about $B = 1.2~\text{T}$, but at $B = 2.6~\text{T}$ a new signal starts appearing as displayed in the upper part of Figure 2a. We attribute this new signal to the anti-crossing between the lowest Zeeman-split quintet states Q$_{2-}$(2,14) and Q$_{2-}$(1,15) with $m_s = -2$ (green lines in Figure 2c), which have four electrons aligned with the external magnetic field, as illustrated in Figure 2b. Consequently, the Q$_{2-}$ state experiences twice the Zeeman splitting in comparison to the T$_-$ and T*$_-$ states, thereby explaining how the quintet state can become a ground state at a sufficiently large magnetic field (Supporting Figure S2c). Just as the T$_-$ state blocks the singlet state at lower fields, the low energy Q$_{2-}$ state now crosses the T$_-$ state as shown with the dashed white line labeled T$_{-}$-Q$_{2-}$ in Figure 2a, causing triplet-quintet spin blockade in the $B=1.2-2.6~\text{T}$ range. We note that the discontinuity of the fitted dashed T$_{-}$-Q$_{2-}$ line at $B=2$~T may be due to changes in charge in the peripheral environment of the DQD between measurements. Increasing the magnetic field beyond $B = 4.4~\text{T}$ results in a six electron X$_{3-}$ septet state with $S=3$ and $m_s=-3$ (top panel of Figure 2b and yellow line in Figure 2c), which experiences thrice the Zeeman splitting compared to the triplet, to move below the Q$_{2-}$ state and become the new ground state (Supporting Figure S2d). This generates yet another region of spin blockade as seen in the upper part of Figure 2a. Magnetospectrocopy measurements with features similar to those presented here were obtained for a neighbouring even-parity ICT with 2 fewer electrons on QD$_{\text{mw}}$ as well as in another similar device (Supporting Information S4).

\begin{figure}
\centering
\includegraphics[scale=1]{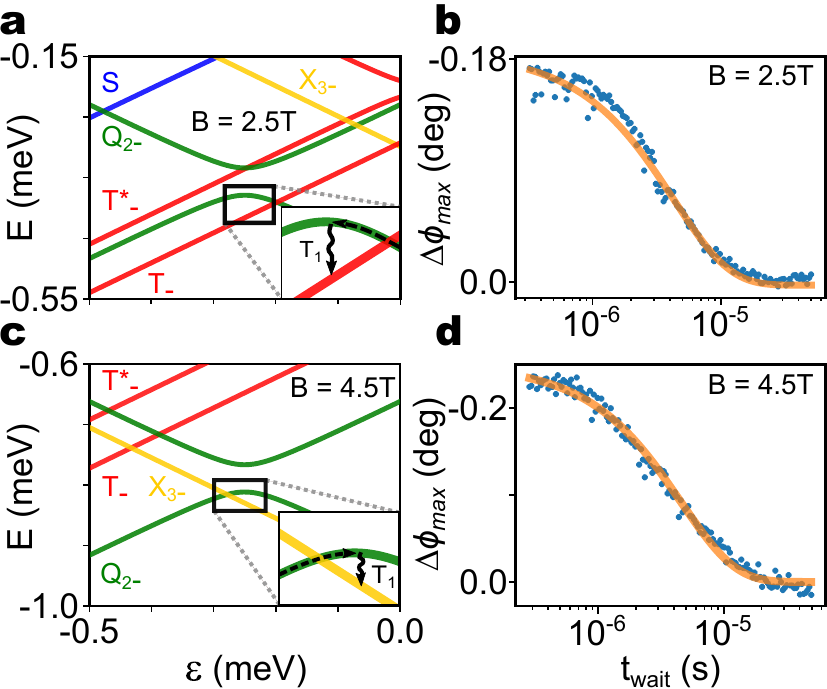}
\caption{Quintet relaxation. \textbf{(a), (c)} Energy levels of the double quantum dots at magnetic fields of $B = 2.5\text{T}$ and $B = 4.5~\text{T}$ simulated based on parameters extracted from magnetospectroscopy measurements presented in Figure 2. The insets highlight the pulsing into the quintet Q$_{2-}$ anti-crossing and the subsequent relaxation into the triplet T$_-$ and septet X$_{3-}$ ground states, respectively. \textbf{(b), (d)} Decay of the maximum ICT phase response $\Delta \phi_{max}$ as a function of wait time $t_{wait}$ at the Q$_{2-}$ anti-crossing. By fitting to a single exponential decay (orange line), we observe quintet relaxation times of $T_{1}^{\text{Q} \rightarrow \text{T}} = 4.26 \pm 0.11 ~\mu \text{s}$ and $T_{1}^{\text{Q} \rightarrow \text{X}} = 4.59 \pm 0.11 ~\mu \text{s}$.}
\label{fig3}
\end{figure}

From the measurements presented in Figure 2a, we compile quantitative information about the energy spectrum of the DQD, which we use to produce Figure 2c showing the singlet (blue), triplet (red), quintet (green) and septet (yellow) states at $B=0~\text{T}$. We obtain $\alpha=0.345$ from the slopes of the asymmetrically vanishing ICT signals (dashed lines in Figure 2a). We get the minimum energy separation between states with same total spin number -- the tunnel couplings -- from the full width at half maximum (FWHM) of the signals at a fixed $B$ (Supporting Information S5) and thus extract $\Delta_\text{S}=80$~$\mu$eV for the singlet at $B=0$~T, $\Delta_\text{Q}=45$~$\mu$eV for the quintet at $B=3.71$~T and estimate $\Delta_\text{T}=35$~$\mu$eV and $\Delta_\text{T*}=20$~$\mu$eV at $B=1.02$~T for the T$_-$(1,15)-T$_-$(2,14) and the T*$_-$(1,15)-T$_-$(2,14) tunnel coupling, respectively. Finally, from the signal position on the $V_\text{Grf}$ axis and the magnetic fields at which the different spin blockades occur, we extract the energy splitting between states in the (2,14) and (1,15) configurations, which we reference for simplicity as (2,0) and (1,1), respectively. We get the singlet-triplet splitting on the (2,0) side, $\delta_\text{ST}^{20}=80$~$\mu$eV, the singlet-triplet$^*$ splitting on the (1,1) side, $\delta_\text{ST*}^{11}=80$~$\mu$eV, the singlet-quintet splitting on the (1,1) side $\delta_\text{SQ}^{11}=360$~$\mu$eV and the (2,0) side $\delta_\text{SQ}^{20}=100$~$\mu$eV and lastly the singlet-septet splitting on the (2,0) side $\delta_\text{SX}^{20}=570$~$\mu$eV. The smallest splittings, $\delta_\text{ST}^{20}$ and $\delta_\text{ST*}^{11}$, may be associated with valley splittings, while the others involve combinations of valley and orbital excitations in QD$_{\text{mw}}$. Using the parameters introduced above, we build the multi-spin Hamiltonian of the system (Supporting Information S6) and extract the eigenenergies, $E_i$, as plotted in Figure 2c. Using the expression for $C_\text{Q}$ and the magnetic field dependence of the eigenenergies, we obtain the simulated magnetospectrocopy map of Figure 2d for which we included a finite electron temperature $T=175~\text{mK}$. The good match between data and simulation confirms our understanding of the DQD energy spectrum and thereby opens up for the possibility of probing the dynamical properties of the high-spin states accessible in this DQD.

Spin relaxation time $T_1$ is an important metric as it ultimately limits coherence and determines the minimum readout time to achieve high fidelity spin readout\cite{kloeffel2013prospects}. The spin relaxation in silicon has been measured for single spins \cite{morello2010single}, singlet-triplet systems \cite{west2019gate}, hybrid qubits \cite{shi2012fast}, and hole spins \cite{crippa2019gate, van2018readout}. In the following we explore spin relaxation from spin quintet states. To determine the relaxation time of the Q$_{2-}$ state, we first set $B = 2.5~\text{T}$ at which point the DQD is under triplet-quintet spin blockade. Figure 3a shows the energy levels at $B = 2.5~\text{T}$ as simulated based on the parameters extracted from magnetospectroscopy. At this field, the Q$_{2-}$ anti-crossing at $\varepsilon \sim -0.25$ meV lies at energies slightly higher than the energy of the T$_-$ state, however at sufficiently large $\varepsilon$, Q$_{2-}$(2,14) is the ground state. To probe the Q$_{2-}$ anti-crossing, we therefore apply a 50\% duty cycle square wave voltage pulse to G$_{\text{mw}}$ that is initialised (I) from the (1,14) configuration and then pulsed to the measurement point (M) on the ICT as seen in the inset of Figure 1b. When crossing the (1,14)-(2,14) charge transition, an electron is loaded into QD$_\text{rf}$ at a rate $\Gamma \gtrsim f_\text{rf}$, given that the transition produces a measurable phase shift in the resonator response~\cite{gonzalez2015probing}. We set the rise/fall time of the pulse to 10~ns, much slower than the tunnelling time, such that the system is initialised in the energetically favoured Q$_{2-}$(2,14) state. The quintet state is subsequently followed through adiabatic passage to the Q$_{2-}$ anticrossing (point M), as illustrated in the inset of Figure 3a, causing a non-thermal occupation probability. Finally, the electron is unloaded and the system reset via the reverse procedure.
 
To measure $T_1$, we wait for a time $t_{wait}$ at point M, during which the Q$_{2-}$ state relaxes to the T$_-$ state with a characteristic relaxation time $T_{1}$ as indicated in the inset of Figure 3a. When spending very little time ($t_{wait} \ll T_1$) at the measurement point of the pulse sequence, the probability that the Q$_{2-}$ state relaxes tends to 0, which leads to lifting of the spin blockade and generation of the ICT phase shift signal from the curvature of the Q$_{2-}$ state according to Equation (1). Contrarily if $t_{wait} \gg T_1$, the probability that the Q$_{2-}$ state relaxes to the non-curving T$_-$ state tends to 1 and we therefore no longer measure the ICT signal. 

We repeat the pulse sequence described above $N>2.4 \cdot 10^5$ times for various $t_{wait}$ and with the rf drive continuously on, thus measuring the average $\Delta \phi$ of the voltage region around the Q$_{2-}$ anti-crossing. From fitting the $\Delta \phi$ line-trace to $\Delta \phi \propto c_1 \left( (\varepsilon-\varepsilon_0)^2+(\Delta_\text{Q})^2 \right)^{-3/2}+c_2$, where $\varepsilon_0$ and $\Delta_\text{Q}$ is the detuning and the tunnel coupling of the quintet anti-crossing, respectively, and $c_i$ are constants \cite{mizuta2017quantum}, the maximum phase shift $\Delta \phi_{max}$ is extracted. In order to obtain the Q$_{2-}$-T$_-$ relaxation time, $\Delta \phi_{max}$ is then plotted as a function of $t_\text{wait}$ and fitted to a single exponential decay, $A_1 e^{-t_{wait}/T_{1}^{\text{Q} \rightarrow \text{T}}}+A_2$, where $A_i$ are constants, as shown in Figure 3b, whereby we find $T_{1}^{\text{Q} \rightarrow \text{T}} = 4.26 \pm 0.11 ~\mu \text{s}$.



Setting the magnetic field to $B=4.5~\text{T}$, we now study the quintet in the region of quintet-septet spin blockade, illustrated in Figure 3c which shows the simulated energy levels at this magnetic field. While similar to the lower field triplet-quintet scenario, here the Q$_{2-}$ state is the ground state at $\varepsilon \lesssim -0.25~\text{meV}$, i.e. in the (1,15) configuration, rather than at $\varepsilon \gtrsim-0.25$ meV. This change affects the pulsing scheme required to determine the relaxation time. The voltage pulses on G$_{\text{mw}}$ are modified accordingly, starting instead from the (2,15) state (I$^\prime$) in order to initialise the system in the energetically favoured Q$_{2-}$(1,15) before moving via adiabatic passage to the measurement point (M$^\prime$) as shown in the insets of Figures 1b and 3c. As in the quintet-triplet case, $\Delta \phi_{max}$ is extracted from the ICT phase response and fitted against a single exponential decay, $A_3 e^{-t_{wait}/T_{1}^{\text{Q} \rightarrow \text{X}}}+A_4$, shown in orange in Figure 3d, which results in a Q$_{2-}$-X$_{3-}$ relaxation time $T_{1}^{\text{Q} \rightarrow \text{X}} = 4.59 \pm 0.11 ~\mu \text{s}$ that is comparable to $T_{1}^{\text{Q} \rightarrow \text{T}}$. 
While there are no other reported quintet relaxation timescales to benchmark these results to, it is about three orders of magnitude smaller than previously reported relaxation times in singlet-triplet DQD systems \cite{west2019gate, pakkiam2018single, prance2012single}. The mechanism explaining the faster quintet relaxation may be subject for further studies.

In summary, we have demonstrated the formation of a tunnel-coupled DQD in the channel of a split-gate silicon transistor. By embedding the device in an LC electrical resonator and performing gate-based dispersive sensing, we have determined the charge state of the DQD down to the last electron. By tuning to the (1,15)-(2,14) interdot charge transition and studying the quantum capacitance of the system as a function of magnetic field, we have found evidence of multi-particle high-spin states not studied before, namely electron spin quintets ($S=2$) and spin septets ($S=3$). From the magnetospectroscopy measurements, we determined the energy spectrum as a function of energy detuning between dots as well as the coupling energy between states with equal spin numbers. Additionally, we have developed a model to describe the magnetic field dependence of the dispersive response based on the multi-state quantum capacitance of the system. The model reproduces the experimental results well and generally provides a tool for understanding dispersive signal in systems with a complex spin configuration. Finally, non-equilibrium studies measured under spin blockade and presented here demonstrate a typical quintet relaxation time, $T_1 \sim 4~\mu \text{s}$, at the hybridisation point. Overall, our results provide a way to reconstruct the energy spectrum of complex spin systems and open up for the possibility of using CMOS quantum dots as a tuneable test bed for studying the interactions and dynamics of high-spin systems. 



This research has received funding from the European Union's Horizon 2020 Research and Innovation Programme under grant agreement No 688539 (http://mos-quito.eu). T.L. acknowledges support from the EPSRC Cambridge NanoDTC, EP/L015978/1. M.F.G.Z. acknowledges support from the Royal Society and the Winton Programme for the Physics of Sustainability. Y.M.N. and J.L. acknowledge support from the French National Research Agency (ANR project MAQSi).


\bibliography{Quintet}
\end{document}